\documentclass[%
 aip,
 jmp,%
 amsmath,amssymb,
 reprint,%
]{revtex4-2}

\usepackage{graphicx}% Include figure files
\usepackage{dcolumn}% Align table columns on decimal point
\usepackage{bm}
\usepackage{cleveref}
\begin{document}

\preprint{APS/123-QED}

 \title[Scattering of classical gravitons with matter fields in the  Null Surface Formulation]{Scattering of classical gravitons with matter fields in the  Null Surface Formulation}

\author{Carlos N. Kozameh}
\email{carlos.kozameh@unc.edu.ar}
\author{Emmanuel A. Tassone}%
 \email{emmanuel.tassone@unc.edu.ar}
\affiliation{FaMAF, Universidad Nacional de C\'ordoba, 5000, C\'ordoba, Argentina}%

\date{\today}% It is always \today, today,

\begin{abstract}

Using a set of field equations in the Null Surface Formulation we obtain the linearized coupling between the gravitational and matter fields. We first derive a formula for the metric of the space time and then we use this formula to study the scattering of incoming gravitational waves when matter is present, obtaining explicit formulae relating the radiation modes at past and future null infinity for a general asymptotically flat spacetime. An example application is made at the end of this work when the matter field is a massless real scalar field. The relevance of this result for a perturbation procedure is discussed.
\end{abstract}

\maketitle

\section{\label{int} Introduction}
The general theory of relativity is a geometric theory, the main field is the metric tensor from which the curvature and other geometrical structures are constructed.\\ The null surface formulation (NSF) is an equivalent formulation of general relativity written in terms of two scalar functions, one giving the conformal structure of the spacetime and the other one providing the conformal factor of the metric tensor. Thus, the main variables of the NSF are these two scalar functions, and the curvature related quantities, such as the metric tensor, are derived concepts.\\
Introducing a bundle of null directions for the points of the spacetime with local coordinates $(x^a,\zeta,\bar{\zeta})$, the two scalars functions are the conformal factor of the theory, $\Omega(x^a,\zeta,\bar{\zeta})$, and a family of characteristic null surfaces of the metric, $Z(x^a,\zeta,\bar{\zeta})$ which satisfies a set of two differential equations, the so-called  "metricity conditions".\\
Mathematically speaking, the ten Einstein equations with components depending of the spacetime points $x^a$ are exchanged by one equation for the scalars $\Omega(x^a,\zeta,\bar{\zeta})$ and $Z(x^a,\zeta,\bar{\zeta})$ along with the two metricity conditions\cite{frittelli1995gr}. The new system of differential equations depends on six variables instead of the four spacetime points $x^a$. This differential system equation continues being nonlinear and thus numerical and perturbative approaches must be employed in order to find solutions.\\
Recently, there have been some further developments of geometrical structures via the NSF, namely, a definition of center of mass and intrinsic angular momentum\cite{kozameh2016center},  a study of global variables in black hole coalescence\cite{tassone2021numerical,tassone2022final}, and the field equations for the classical graviton were developed\cite{bordcoch2016asymptotic}. The classical graviton is a Ricci flat spacetime that represents the self interaction of incoming gravitational waves. In addition, a perturbation procedure to obtain the solution was given and at each level of this perturbation scheme  the constructed solution was finite. This approach could be of importance when dealing with quantization of pure gravity and hints of this quantization were given before the perturbation scheme was implemented. \cite{frittelli1997fuzzy,frittelli1997quantization}

It is also important to extend this approach to the interaction of classical gravitons with other fields but a word of caution must be given. The formalism is not adapted to study asymptotically flat spacetimes with singularities since an ab initio assumption is broken, namely, that it is possible to find regular null cone cuts at future or past null infinity. Thus, the scenario we envisage is of a stress energy tensor constructed from suitable fields that preserve the regularity of the null cone cuts, as for example the interaction of gravitons with photons.

In this work we use the NSF to study the interaction between matter and the gravitational field in the asymptotic limit and an explicit relationship between the incoming and outgoing gravitational fields is derived. We thus generalize the concept of classical gravitons previously given. The gravitational radiation field at future and past null boundaries possess the same mathematical structure as in the vacuum case\cite{bordcoch2016asymptotic} but now there is an interaction term inside the spacetime. The results are equivalent to the classical scattering of fields at the first non trivial level, usually called tree level, where the first order interaction term between the interacting fields is used to compute the incoming and outgoing fields.\\ 
The paper is organized as follows. In Sec.\ref{sectionII} we give an outline of the NSF and particularly, we explain the linearized version of the formulation which is simpler mathematically and allows us to study the interaction of the fields analytically. Then, we derive the NSF equation including matter. In Sec.\ref{sectionIII}, we derive the relation between the metric and the NSF variables and we also analyse the ingoing and outgoing characteristic null surfaces to linear order. In case of no matter, we relate the modes in the decomposition of gravitational radiation. We then perform a similar calculation incorporating the matter field tensor and, in particular, we study the scattering of gravitons with scalar waves, showing that we obtain an expected result at the tree level approximation. Finally, we finish the work with a summary and conclusions in Sec.\ref{sectionVI}.

% Introduction to Linear NSF
\section{\label{sectionII}The linearized Null Surface Formulation}
The Null Surface Formulation recasts general relativity as a theory of null surfaces interacting with matter via the field equations. In NSF the Lorentzian metric of the spacetime $g_{ab}$ is constructed from a conformally invariant part $h_{ab}$ and a conformal factor $\Omega$. The conformal metric is obtained from knowledge of special null surfaces obtained from a real function $Z$ given on the sphere of null directions bundle. Given $Z=Z(x^a,\zeta,\bar{\zeta})$, with $x^a$ points of the spacetime and $(\zeta,\bar{\zeta})$ stereographic coordinates on the sphere of null directions, $Z=const.$ is a null surface on the spacetime. From knowledge of $Z$ one then constructs a conformal metric $h_{ab}[Z]$. In addition, one gives a second real function $\Omega=\Omega(x^a,\zeta,\bar{\zeta})$, a conformal factor, that plays a dual role: it is used to obtain a conformal metric that only depends on the spacetime points and it yields an Einstein metric via the field equations.  From knowledge of these two functions one obtains $$g_{ab}(x^a)=\Omega^2 h_{ab}[Z].$$ Since the equation $Z= const.$ defines a null hypersurface, $Z$ must satisfy
\begin{equation}\label{g00}
g^{ab}\partial_a Z\partial_b Z=0.
\end{equation}
It is clear from (\ref{g00}) that the conformal structure does not depend on $\Omega$. It also follows from this equation that the null vector $Z^a=g^{ab}\partial_b Z,$ satisfies the homogeneous geodesic equation, thus defining an affine length $s$. 

Directly from (\ref{g00}), and taking $\eth$ and $\bar{\eth}$ derivatives \cite{frittelli1995gr}, one obtains the components of the conformal metric. The non trivial coefficients of $h^{ab}$ are functions of a single scalar $\Lambda$ defined as $\Lambda = \eth^2 Z$. Once the conformal metric coefficients have been obtained one finds a condition on $\Lambda$, namely

$$
    \eth^3 (g^{ab}\partial_a Z\partial_b Z)=0 \implies g^{ab}(3 \partial_a \eth Z\partial_b \Lambda + \partial_a Z \partial_b \eth \Lambda)=0.
$$

This condition can be rewritten as 
\begin{equation}\label{Wunchsmann}
\frac{\partial \eth \Lambda}{\partial s}+3 \eth Z^b\partial_b \Lambda=0.
\end{equation}

Only for functions $\Lambda$ that satisfy condition (\ref{Wunchsmann}) it is possible to obtain a conformal metric. In what follows we assume condition (\ref{Wunchsmann}) is satisfied.\\
One can also show that directly from

$$
    \eth^2 \bar{\eth}^2(g^{ab}\partial_a Z\partial_b Z)=0,
$$
one obtains a relationship between $\Omega$ and $\Lambda$, namely,
\begin{equation}\label{g01}
\frac{\partial \bar{\eth}^2 \Lambda}{\partial s}= \eth \bar{\eth}(\Omega^2)+g^{ab}\partial_a \Lambda\partial_b \bar{\Lambda}.
\end{equation}

which can be formally integrated giving \cite{bordcoch2016asymptotic},
\begin{equation}\label{Lambda}
\bar{\eth}^2 \Lambda= \eth^2 \bar{\sigma}(Z,\zeta,\bar{\zeta})+ \bar{\eth}^2 \sigma(Z,\zeta,\bar{\zeta})+\int _Z^\infty \dot{\sigma}\dot{\bar{\sigma}} du-\int _s^\infty(\eth \bar{\eth}(\Omega^2)+g^{ab}\partial_a \Lambda\partial_b \bar{\Lambda})ds',
\end{equation}
where $\sigma(u,\zeta,\bar{\zeta})$, the Bondi shear, is directly related to the gravitational radiation reaching null infinity and $\dot{\sigma}$ is the derivative with respect to the Bondi time $u$. It is worth mentioning that conditions (\ref{Wunchsmann}) and (\ref{g01}) have been used to obtain the free data at null infinity\cite{bordcoch2016asymptotic}, otherwise one should have more general free function when (\ref{g01}) is integrated along a null geodesic from the point $x^a$ to null infinity. An explicit calculation for the linearized approximation given in \cite{frittelli1995dynamics} serves to illuminate this point. \\

In the NSF approach the function $\Lambda$ from \cref{Lambda} plays an important role since the conformal metric is completely given in terms of this function and its vanishing yields a flat conformal metric. Thus, one can implement a perturbation procedure directly from knowledge of $\Lambda$ and write down the lowest non trivial formulation from a linearized approximation. 

The conformal factor is fixed via the field equations. Directly from the conformal relationship between the two metrics one finds\cite{wald2010general}:
\begin{align}\label{EinsteinNSF}
    2\frac{\partial^2\Omega}{\partial s^2}  = Z^a Z^b( R_{ab}[h]- R_{ab}[g])\Omega.
\end{align}
 The scalar $R_{ab}[h]Z^a Z^b$ is a quadratic function of $\Lambda$ and it thus vanishes at a linearized approximation, whereas $R_{ab}[g]$ can be replaced by the trace free part of the energy-momentum tensor $T_{ab}$. 
The three scalar equations (\ref{Wunchsmann}), (\ref{g01}) and (\ref{EinsteinNSF}) are completely equivalent to the full Einstein equations for a metric $g_{ab}$. The linearized version of \cref{Lambda,EinsteinNSF} is obtained by first giving the zeroth-order solution that yields a flat metric, namely,

\begin{align} \label{FlatFoliation}
Z_0 = x^a\ell_a,  \; \; \Omega_0= 1.
\end{align}
with $\ell^a$ a null vector defined as $\ell^a=\frac{1}{\sqrt{2}}(1,r^i)$ and $r^i$ the unit vector on the sphere of null directions, and $x^a $ a point in the flat spacetime.\\
One then writes down a linearized departure from the zeroth order solution as
\begin{align}
Z = x^a\ell_a + Z_1, \; \; \Omega = 1+\Omega_1,
\end{align}
and its corresponding equations of motion
\begin{align}\label{Eq8}
    \frac{\partial (\bar{\eth}^2 \eth^2 Z_1)}{\partial s}= 2\eth \bar{\eth}(\Omega_1),
\end{align}
\begin{align}\label{EinsteinNSF2}
    2\frac{\partial^2\Omega_1}{\partial s^2}  = -T_{ab}Z^a Z^b,
\end{align}
which can be integrated following the null geodesic either to future or past null infinity to obtain the advanced or retarded solutions. Note that $\Omega_1$ only depends on the stress energy tensor and vanishes in absence of matter fields. Assuming we are solving \cref{Eq8,EinsteinNSF2}  for an asymptotically flat spacetime, we can integrate \cref{Eq8} and obtain

\begin{align}\label{FinalNSF}
&\bar{\eth}^2\eth^2 Z_1 =\bar{\eth}^2 \sigma(Z_1,\zeta,\bar{\zeta}) + \eth^2 \bar{\sigma}(Z_1,\zeta,\bar{\zeta}) -2\int _0^\infty \eth \bar{\eth}(\Omega_1)ds +\mathcal{O}(\Lambda^2).
\end{align}

The function $Z_1$, on the other hand, depends both on the matter fields as well as $\sigma$. Note also that the r.h.s. of \cref{FinalNSF} contains non-linear expressions of $Z_1$ since the Bondi shear is an arbitrary function at null infinity. Thus one must linearized the above equation using a perturbation procedure. Eq. (\ref{FinalNSF}) is a non-homogeneous 4th order elliptic equation on the sphere and its solution can be found by convoluting the inhomogeneity with the corresponding Green function \cite{ivancovich1989green}. Note that the above \cref{FinalNSF} given in this work differs from a previous derivation in \cite{frittelli1995linearized}. The calculation showed in the latter reference is more involved, as one can see from comparing the terms containing the conformal factor as well as the extra integral along the Bondi time that is not present in our derivation.  The main reason for the difference between these equations is that the one given on the previous work has been derived combining the two metricity conditions. We have used our
derivation for simplicity.  It is worth pointing out this difference since we heavily use eq. (\ref{FinalNSF}) in the results presented in this work.

Higher order terms on a perturbation scheme can be implemented in \cref{Lambda} by making a perturbation series in the null foliation

\begin{align}\label{Zperturbation}
    Z=Z_0+Z_1+Z_2+...
\end{align}

where $Z_0$ correspond to a null foliation in Minkowski space as in \cref{FlatFoliation}, and $Z_1$ satisfies equations (\ref{EinsteinNSF2}) and (\ref{FinalNSF}). This perturbation procedure is carried on giving the $n-1$ and $n-2$  orders that were previously obtained to solve for  $Z_n$. At any order of the perturbation a conformal structure is derived and the new solution is a field on this spacetime.

We will show later in Sec.\ref{sectionIII} and \ref{sectionIV} this scheme to be useful when studying gravitational radiation and its interaction with the energy-momentum tensor. The linearized NSF allows to express $\Omega(x^a ,\zeta,\bar{\zeta})$ as an explicit functional of the energy momentum tensor leaving the theory with just one main scalar, $Z$, which can be obtained by iterating to higher orders.

% Agregar comentario sobre los términos nuevos que falta de g01
% Construction for doing scattering
\section{Scattering in the linearized NSF}\label{sectionIII}
In this Section, we derive the relation between the incoming and outgoing radiation at the null boundary of a general asymptotic spacetime, solution to \cref{FinalNSF}. With this end, we need to find a point in the spacetime where both kind of radiations are well defined, relate this point to null infinity through geodesic paths and, finally, apply the uniqueness of the metric to the corresponding expansion order. Thus, we will use the subscripts $+$ and $-$ to denote quantities associated to future and past null infinity respectively. In this sense, $\sigma_+$ will stand for outgoing radiation at future null infinity and $\sigma_-$ for ingoing radiation coming from past null infinity. \\
The reader should remind that the linearized \cref{FinalNSF} still remains nonlinear in the variable $Z$, as the dependence of the shear $\sigma(Z,\zeta,\bar{\zeta})$ on $Z$ is not known. To tackle this problem, we use the perturbative approach from \cref{Zperturbation} at the end of this section.

% Tensorial spin-s harmonics introduction

\subsection{Tensorial spin-$s$ harmonics}
When working with gravitational radiation (or any problem involving spin-$s$ fields) it is common the use of spin-$s$ spherical harmonics $Y^s_{lm}$ to expand the angular dependence of functions that are defined at null infinity. In this work we'll use instead an equivalent complete base of orthogonal functions known as tensorial spin-s harmonics.\\

Given a Newman-Penrose (NP) null tetrad $(l^a ,n^a ,m^a ,\bar{m}^a )$ for a flat metric, we can construct a 3-dimensional orthonormal base $(c^i , m^i ,\bar{m}^i)$ for any tangent space of a point $p$ in $S^2$. This 3-dimensional base can be constructed as follows: First, we construct the timelike and spacelike vectors $t^a=l^a+n^a$ and $c^a=l^a-n^a$. Then, we take the projections of our 4-dimensional base normal to the vector $t^a$. The three obtained independent projections will be called $(c^i , m^i ,\bar{m}^i)$. These Euclidean vectors are the main ingredient in the construction of the tensor spin-$s$ harmonics. Moreover, the three dimensional Euclidean base can be used to set up a general $l$-dimensional tensor base by doing the exterior product of the vectors $c^i ,m^i ,\bar{m}^i $. This concept is what the tensor spin-$s$ harmonics embrace. We mention in this subsection only a few properties of the construction but the reader may find more in the following Ref. \cite{newman2005tensorial,mandrilli2020correspondence}.\\
The tensorial spin-$l$ harmonics, denoted $Y^l_{lI_l}$ with $I_l$ a set of $l$ indices, are defined as the $l$-times product of the space-like vector $m^i$. Then, the complex covariant derivative on the manifold defined in $S^2$, known as $\eth$, along with its conjugate $\bar{\eth}$ behave as "ladder operators" with respect to the spin index (the number in the superscript). These ladder operators allow to obtain the whole set of representations of the tensors $Y^s_{lI_l}$. A relevant property from these tensors is

\begin{align*}
    \bar{Y}^s_{lI_l}=Y^{-s}_{lI_l}.
\end{align*}

% Antipodal transformations
\subsection{Antipodal transformations on the sphere}\label{AntipodalTransformations}
To link incoming and outgoing radiation at null infinity, we need to introduce the notion of antipodal points on the sphere. The antipodal points are those diametrically opposite to each other in the sphere. Then, we define the antipodal transformation to be the one that carries a point of the sphere to its antipodal point.\\
In the usual spherical chart $(\theta,\phi)$, the antipodal transformation reads

\begin{align*}
    (\theta,\phi) \rightarrow (\pi-\theta,\pi+\phi) ,
\end{align*}

or in stereographic coordinates

\begin{align*}
    (\zeta,\bar{\zeta}) \rightarrow (-1/\bar{\zeta},-1/\zeta).
\end{align*}

We denote the antipodal transformation with the symbol $\widehat{}$ , i.e, $\widehat{\zeta}=- 1/\bar{\zeta}$. Given a tensorial spin-$s$ harmonic $Y^s_{lI_l}(\zeta,\bar{\zeta})$, we have

\begin{align}
    Y^s_{lI_l}(\widehat{\zeta},\widehat{\bar{\zeta}})=(-1)^{l}Y^{-s}_{lI_l}(\zeta,\bar{\zeta})
\end{align}. 

In particular, if we write $l_-^a=\frac{1}{\sqrt{2}}(-1,r^i)$ with $r^i$ the corresponding spatial vector, the antipodal transformation is

\begin{align}\label{antipodall}
    \widehat{l}_-^a=\frac{1}{\sqrt{2}}(-1,\widehat{r}^i)=\frac{1}{\sqrt{2}}(-1,-r^i)=-\frac{1}{\sqrt{2}}(1,r^i)=-l_+^a
\end{align}

\cref{antipodall} is the antipodal transformation applied to a null vector $l_-^a$ defined at past null infinity. Hence, this vector has its antipodal point at minus the position of a vector defined at future null infinity.\\
The antipodal transformation on the derivative operator $\eth$, which will appear later in the calculations, can be proven to be
$$
\widehat{\eth}=-\bar{\eth}\ .
$$

% Relation between metric and foliation at first order
\subsection{Relation between null cuts at $\mathcal{I}^+$ and $\mathcal{I}^-$}
As discussed in Sec.\ref{sectionII}, the condition of the null cone cuts to be null on a general asymptotic spacetime 
$$
g^{ab}\partial_a Z \partial_b Z=0
$$
can be written for each perturbation order from \cref{Zperturbation}. Thus, the n-order null condition reads

\begin{align}\label{NullConditionPerturbed}
\sum_{n}^{\infty} \sum_{r+s=0}^{n} g_{n-r-s}^{a b} \partial_{a} Z_{r} \partial_{b} Z_{s}=0    
\end{align}
with,
$$g_0^{ab}=\eta^{ab},$$
and,
$$g_1^{ab}=2 \Omega_1 \eta^{ab} + h_1^{ab},$$
the first order correction to the flat metric. 

The zero order null condition is just the vector $l^a$ to be null $\eta^{ab}l_a l_b=0$. The first non-trivial null condition appears at order $n=1$,

\begin{align}\label{FirstOrderNullCondition}
    h^{a b} l_{a} l_{b}+2 \eta^{a b} l_{a} \partial_{b} Z=0 ,
\end{align}
where we have dropped the subindex $1$ in $h_1^{a b}$ and $Z_1$ for simplicity.
Up to first order, we can use the Minkowski metric to raise or lower indices, thus, $\eta^{ab}l_a=l^b$ and

\begin{align}\label{FirstOrderNullCondition2}
    h_{a b}(x^a ,\zeta) l^{a}l^{b} =-2 l^{a} \partial_a Z(x^a ,\zeta)\ .
\end{align}

\cref{FirstOrderNullCondition2} relates the first order correction $h_{ab}$ with the first order correction of the null foliation $Z$. Note that $Z$ plays the role of a potential for $h_{ab}$. As we will see below the solution to the linearized equation for $Z$ shows its explicit dependence on the gravitational radiation $\sigma$ and the stress energy tensor $T_{ab}$, and it is completely equivalent to a standard formulation using an advanced or retarded Green function.\\
Eq.(\ref{NullConditionPerturbed}) can be continued to higher orders to obtain nonlinear terms. We will only keep the first non trivial terms in this equation, since the purpose of this work is to obtain the tree diagrams arising in a scattering formulation of gravitational radiation interacting with matter fields.\\

% Scattering flat space
%%%%%%%%%%%%%%%%%%%%%%%
\subsection{Solution in absence of matter}
In case of no matter fields, i.e $T^{ab}=0$, we have $\Omega_1=0$ in \cref{FinalNSF}. Considering this simplification, we want to find the first order correction to the null flat foliation $Z_0$. Hence, we need to solve \cref{FinalNSF} to first order 

\begin{align}\label{FinalNSFfirstorder}
&\bar{\eth}^2\eth^2 Z(x^a ,\zeta,\bar{\zeta})=\bar{\eth}^2 \sigma(Z_0,\zeta,\bar{\zeta}) + \eth^2 \bar{\sigma}(Z_0,\zeta,\bar{\zeta}) .
\end{align}
 
 The solution can be found by convoluting the inhomogeneity of the equation with the Green function on the sphere \cite{ivancovich1989green},
 
\begin{align}\label{G00'}
 G_{00'}(\zeta,\bar{\zeta},\zeta',\bar{\zeta'})= \frac{1}{4\pi} l^{+a} l^{\prime +}_a ln(l^{+a} l^{\prime +}_a).
 \end{align}
 
 The solution reads
 
 \begin{align}\label{Z1+}
     Z^{+}(x^a,\zeta,\bar{\zeta})= \oint_{S^2} G_{00'}\left(\bar{\eth}^{\prime 2} \sigma^{+}(x^al^{\prime +}_a,\zeta',\bar{\zeta'})+\eth^{\prime 2} \bar{\sigma}^{+}(x^al^{\prime +}_a,\zeta',\bar{\zeta'})\right) dS^{\prime},
 \end{align}
where we have used that $Z_0^{+}=u=x^a l^{+}_a$.
The superscript in $Z^+$ is used to denote that the future null directed foliation has been used to construct the future null cone cut $x^a l^{+}_a$. Similarly, $\sigma^+$ denotes the outgoing radiation at future null infinity.\\
Another solution to \cref{FinalNSFfirstorder} can be found in terms of ingoing radiation
 
 \begin{align}\label{Z1-}
     Z^{-}(x^a ,\zeta,\bar{\zeta})= \oint_{S^2} G_{00'}\left(\bar{\eth}^{\prime 2} \sigma^{-}(-x^a\ell^{\prime-}_{a},\zeta',\bar{\zeta'})+\eth^{\prime 2} \bar{\sigma}^{-}(-x^a\ell^{\prime -}_{a},\zeta',\bar{\zeta'})\right) dS^{\prime}.
 \end{align}
where we have used the null cone cut $Z^-_0=-x^a\ell^-_{a}$,  given by the intersection of the past null cone from $x^a$ with past null infinity. Note also that at past null infinity the cut is described by  $v=Z^-_0$, with $v$ the usual advanced time coordinate.\\
Given a point in the spacetime $x^a $ and the metric $g^{ab}(x^a )$ at that point, we can write the metric in terms of the ingoing ($\sigma^-$) or outgoing radiation ($\sigma^+)$ using the scalar field $Z$.
On the other hand, from the uniqueness of the metric tensor $g^{ab}(x^c)$ both descriptions of the metric must coincide, i.e $g^{ab}_-(x^c )=g^{ab}_+(x^c )$. Thus, one should have,

\begin{align}
    h^{+}_{1ab}(x^c)l^{+a}l^{+b}=h^{-}_{1ab}(x^c)\hat{l}^{-a}\hat{l}^{-b}.
\end{align}

In absence of a stress energy tensor the first order deviation $h_{1}^{ab}$ only depends on $x^a$.
Using \cref{FirstOrderNullCondition2} we find a relation between first order outgoing and ingoing null surfaces

\begin{align}\label{ScatteringRelation}
   l^{\prime a} \partial_a Z^{+}=\hat{l}^{+a}\partial_a\widehat{Z^{-}}
\end{align}
which can also be written as
\begin{align}\label{ScatteringRelationB}
    l^{+a} \partial_a (Z^+ +\widehat{Z^-})=0.
\end{align}

Eq. (\ref{ScatteringRelationB}) can be thought of as an equation for two different functions of the same variables.  Taking into account that $l^{-a}=-\hat{l}^{+a}$, $G_{00'}(\zeta,\zeta')=G_{00'}(\hat{\zeta},\hat{\zeta}')$ and performing a change of variables  $\zeta \to \hat{\zeta}$, one rewrites $Z^{-}(x^a ,\hat{\zeta},\hat{\bar{\zeta}})$ in  \cref{ScatteringRelationB} as

\begin{align}\label{Zminus_antipodal}
     Z^{-}(x^a ,\hat{\zeta},\hat{\bar{\zeta}})= \oint_{S^2} \left(\eth^{\prime 2}G_{00'} 
     \sigma^{-}(x^a\ell^{\prime +}_{a},\widehat{\zeta'},\widehat{\bar{\zeta'}})+\bar{\eth}^{\prime 2} G_{00'}\bar{\sigma}^{-}(x^a\ell^{+'}_{a},\widehat{\zeta'},\widehat{\bar{\zeta'}})\right) dS^{\prime}, 
 \end{align}
where the function $\sigma^{-}(x^a\ell^+_{a},\widehat{\zeta'},\widehat{\bar{\zeta'}})$ is now thought as a function given on future null infinity. A detailed analysis taking into account the parity of this function when expanding in spherical harmonics is given below. From this point of view, the action of $l^{+a} \partial_a$ on $(Z^+ +\widehat{Z^-})$ is to add a common factor $l^{+a}l^{\prime +}_{a}$ to the whole equation. For example, its action on $Z^{-}(x^a ,\hat{\zeta},\hat{\bar{\zeta}})$ yields
$$
\oint_{S^2} l^{+a}l^{\prime +}_{a}\left(\eth^{\prime 2}G_{00'} 
     \dot{\sigma}^{-}(u',\widehat{\zeta'},\widehat{\bar{\zeta'}})+\bar{\eth}^{\prime 2} G_{00'}\dot{\bar{\sigma}}^{-}(u',\widehat{\zeta'},\widehat{\bar{\zeta'}})\right)dS^{\prime},
$$
with $u'=x^a\ell^{\prime +}_{a}$ and now the radiation scalars are derived with respect to this variable. Moreover, using the properties of the spin weighted spherical harmonics, the antipodal transformation of each eigenfunctions is related to its complex conjugate. Thus, we expect a relationship between the Bondi shear at past null infinity and its complex conjugate at future null infinity. An identical result can be obtained by simply demanding that
\begin{align}\label{ScatteringRelationC}
    Z_{1}^++\widehat{Z_{1}^-}=0.
\end{align}
In this case an analogous formula is obtained for $\sigma^{-}(u,\widehat{\zeta},\widehat{\bar{\zeta}})$ instead of $\dot{\sigma}^{-}(u,\widehat{\zeta},\widehat{\bar{\zeta}})$. The potential gauge supertraslation freedom has been fixed by demanding that in absence of radiation the null cone cut is given by $u=x^a\ell^+_{a}$. Note that the physical variables are gauge independent since the metric depends on $\dot{\sigma}$.

%------Modes Relation-------------
\subsection{Modes relation}
To study the relation between the different modes of gravitational radiation, we expand the dependence on the stereographic coordinates from all the physical scalar quantities. That is to say, given a quantity $\eta (u,\zeta,\bar{\zeta})$ with spin weight s, we have
$$
\eta (u,\zeta,\bar{\zeta}) =\sum_l \eta (u)^{I_l}Y^s_{l,I_l}(\zeta,\bar{\zeta}),
$$
where $s=2$ if $\eta=\sigma^0$.\\
Also, the Green function (\ref{G00'}) for the Laplacian on $S^2$ can be expanded as

\begin{align*}\label{GreenFunction}
    G_{00^{\prime}}= \sum_{l=2}^{\infty} \frac{4 \pi}{2l+1} Y_{l,I_L}^{0}(\zeta,\bar{\zeta}) Y_{l,I_L}^{\prime 0}(\zeta',\bar{\zeta}').
\end{align*}

Replacing the above formula in \cref{Z1+} and using Stokes theorem we get the $l,I_L$ contribution to $Z$, i.e.,
\begin{equation}\label{partialZ1plus}
Z^{+}_{l,I_l}=  \oint  \big(Y_{l,I_l}^{-2}  \sigma^{+}(u,\zeta,\bar{\zeta}) +Y_{l,I_l}^{2} \bar{\sigma}^{+}(u,\zeta,\bar{\zeta}) \big) d^2S,
\end{equation}

where the factor $\frac{4\pi}{2l+1}$ has been absorbed in the definition of $Z^{+}_{l,I_l}$.
A similar calculation can be carried out to obtain the right hand side in \cref{ScatteringRelation} but we must also apply the antipodal transformations from Sec. \ref{AntipodalTransformations}, then

\begin{align}\label{partialZ1minus}
         &\widehat{Z}^{-}_{l,I_l}= (-1)^l\oint  \big(Y_{l,I_l}^{-2}  \sigma^{-}(v,\zeta,\bar{\zeta}) +Y_{l,I_l}^{2} \bar{\sigma}^{-}(v,\zeta,\bar{\zeta}) \big) d^2S
\end{align}

To obtain the relation between the different modes from outgoing radiation $\sigma^+_{l,I_L}$ and incoming radiation $\sigma^-_{l,I_l}$, one does the harmonic decomposition of both functions and makes a change of variables as in \cref{antipodall}. The details of the derivation are shown in Appx. \ref{AppendixA}. The coefficients of the ingoing and outgoing radiation are given by the following relations

\begin{align}\label{CoeffRelations}
&\sigma^+_{l,I_l}(w)+(-1)^{l}\bar{\sigma}^-_{l,I_l}(w)=0,
\end{align}
using a positive frequency $w$ and integrating on $S^2$ or,
\begin{align}
&\sigma^+(u, \zeta,\bar{\zeta})+\bar{\sigma}^-(u, \widehat{\zeta},\widehat{\bar{\zeta}})=0,
\end{align}
as a function of the Bondi time $u$ and the stereographic coordinates on the sphere $(\zeta,\bar{\zeta})$. The same result could have been obtained by grouping terms under the same value of $\eth^{\prime 2}G_{00'}$ and c.c. in \cref{ScatteringRelationC}. The alternative derivation given above gives the same result and it can be extended to the case with a non trivial energy momentum tensor.

% Introduccion explicando como encontrar el g01
\subsection{\label{Scattering}Scattering in presence of matter}\label{sectionIV}
When a matter field is present in the spacetime, there is an additional term affecting the form of the foliation $Z$ in \cref{FinalNSF}. The presence of the matter term yields an additional complication since the calculations are technically more involved.  Nevertheless, it is possible to derive an explicit formula for each mode.

The future directed solution of the  NSF equation can be written as

\begin{align}\label{Z1plus}
& Z^{+}_{l,I_l}=  \oint  \big[Y_{l,I_l}^{-2}  \sigma^{+}(u,\zeta,\bar{\zeta}) + c.c. - 2  Y_{l,I_l}^{0}\int_0^\infty ds \bar{\eth} \eth \Omega_1(y^c(s),\zeta,\bar{\zeta})\big] d^2S,
\end{align}
with $y^c(s)= x^c+s l^{+c}$. A geometrical formula can be written by defining $N^+_x$ as the future null cone coming out from the point $x^c$, and $C^+_x$ as the intersection of $N^+_x$ with future null infinity. With these definitions, we have

\begin{align}\label{Z1plusb}
& Z^{+}_{l,I_l}=  \oint_{C^+_x} d^2S \big(Y_{l,I_l}^{-2}  \sigma^{+}(u,\zeta,\bar{\zeta}) + c.c. \big) -2 \int_{N^+_x} Y_{l,I_l}^{0} \bar{\eth} \eth \Omega_1(x^c,s,\zeta,\bar{\zeta})dsd^2S .
\end{align}

The corresponding formulae for $Z^{-}_{l,I_l}$ are
\begin{align}\label{Zminus}
& Z^{-}_{l,I_l}=  \oint  \big[Y_{l,I_l}^{-2}  \sigma^{-}(v,\zeta,\bar{\zeta}) + c.c. - 2 Y_{l,I_l}^{0}\int_0^\infty ds \bar{\eth} \eth \Omega_1(y^c,\zeta,\bar{\zeta}) \big]d^2S,
\end{align}
and 
\begin{align}\label{Zminusb}
& Z^{-}_{l,I_l}=  \oint_{C^-_x} d^2S \big(Y_{l,I_l}^{-2}  \sigma^{+}(v,\zeta,\bar{\zeta}) + c.c.\big) - 2 \int_{N^-_x}  Y_{l,I_l}^{0} \bar{\eth} \eth \Omega_1(x^c,s,\zeta,\bar{\zeta})dsd^2S.
\end{align}
\subsection{The conformal factor in the linearized NSF}
The relation between the Ricci tensor field $R_{ab}$ for a general asymptotic spacetime and its conformal version $\tilde{R}_{ab}$ is given by \cite{wald2010general}

\begin{align}
R_{a b}=\tilde{R}_{a b}+2 \Omega^{-1} \tilde{\nabla}_{a} \tilde{\nabla}_{b} \Omega+\tilde{g}_{a b} \tilde{g}^{c d}\left(\Omega^{-1} \tilde{\nabla}_{c} \tilde{\nabla}_{d} \Omega-3 \Omega^{-2} \tilde{\nabla}_{c} \Omega \tilde{\nabla}_{d} \Omega\right).
\end{align}

As we defined in Sec.\ref{sectionII}, $\Omega$ is the conformal factor of the metric and one of the main scalars from NSF. For the linearized version of NSF, we can expand $\Omega = 1+ \Omega_1$ and $g_{01}=1+2\Omega_1(x^a,\zeta,\bar{\zeta})$. Recalling $\tilde{R}_{ab}=\mathcal{O}(\Lambda^2)$ and contracting with the null tensor $l^a l^b$, we get  

\begin{equation}\label{R00}
R_{a b}l^a l^b=2 l^a l^b\partial_{a} \partial_{b} \Omega_1\ ,
\end{equation}

where we have replaced the covariant by the  partial derivative, valid at this level of approximation. We then use the Fourier version of \cref{R00}

\begin{align}
\Omega_1(k^a,\zeta,\bar{\zeta}) &= \frac{l^a l^b R_{ab}(k^c)}{2(k^c l_c)^2 } \\
&=\frac{T_{ab}(k)l^a l^b}{2(k^c l_c)^2 }\label{deltaOmega},
\end{align}

where $k^a$ is the 4-vector from the transformation and the last equality follows from the fact that $l_a$ is the first vector in the NP null base. Replacing \cref{deltaOmega} in \cref{FinalNSF}, we obtain

\begin{align*}\label{ethbarethg01}
\bar{\eth} \eth \Omega_1(x^c,\zeta,\bar{\zeta})&=  \int d^4k \theta(k^o)[T_{ab}(k^c) e^{-i x^c k_c} + \bar{T}_{ab}(k^c) e^{i x^c k_c}] \bar{\eth} \eth\left(\frac{l^al^b}{(k^cl_c)^2} \right),  
\end{align*}
where $\theta(k^o)$ is the step function.

Hence, each advanced mode  solution to  the NSF equation in presence of matter reads
\begin{equation}\label{Z+WithMatter}
    Z^+_{l,I_l}=\oint   \bigg(Y_{l,I_l}^{-2}\sigma^+(x^c \ell^+_c,\zeta,\bar{\zeta})+ i Y_{l,I_l}^{0}\int d^3k \bar{\eth} \eth\left(\frac{T_{ab}(k^c)\ell^{+a}\ell^{+b}}{(k^c\ell^+_c)^2}\right)\frac{e^{-i x^c k_c}}{k^c\ell^+_c} + c.c. \bigg)d^2S.
\end{equation}
The positive frequency component of this equation is given by
\begin{equation}\label{Z+}
    Z^+(w)_{l,I_l}=\sigma(w)_{l,I_l} + i \int_0^{\infty}dk^o \oint d^2\hat{k} T_{ab}(k^o,w\hat{k}^i) \oint Y_{l,I_l}^{0} \bar{\eth} \eth\left(\frac{l^al^b}{(k^cl_c)^2}\right)\frac{1}{k^cl_c},
\end{equation}
with $k^a=(k^o,w\hat{k}^i)$.

Likewise, the retarded solution for $Z$ yields
\begin{equation}\label{Z-WithMatter}
    Z^-_{l,I_l}=(-1)^l\oint   \bigg(Y_{l,I_l}^{-2}\sigma^-(-x^c \ell^-_c,\zeta,\bar{\zeta}) - i Y_{l,I_l}^{0}\int d^3k \bar{\eth} \eth\left(\frac{\bar{T}_{ab}(k^c)\ell^{+a}\ell^{-b}}{(k^c\ell^-_c)^2}\right)\frac{e^{-i x^c k_c}}{k^c\ell^-_c} + c.c. \bigg)d^2S,
\end{equation}
and
\begin{equation}\label{Z-}
    Z^-(w)_{l,J_l}=(-1)^l\bigg(\sigma^-(w)_{l,J_l} - i \int_0^{\infty}dk^o \oint d^2\hat{k} \bar{T}_{ab}(k^o,w\hat{k}^i) \oint Y_{l,J_l}^{0} \bar{\eth} \eth\left(\frac{l^{-a}l^{-b}}{(k^cl^-_c)^2}\right)\frac{1}{k^cl^-_c}\bigg),
\end{equation}

 Finally (see Appendix), from $Z^{+}(x^a ,\zeta,\bar{\zeta})+Z^{-}(x^a ,\widehat{\zeta},\widehat{\bar{\zeta}})=0$, we get

\begin{equation}\label{Zfinal}
  \sigma^+_{l,I_l}(w) + (-1)^l\bar{\sigma}^{-}_{l,I_l}(w) = i \oint d^2S Y_{l,I_l}^{0}\int_0^{\infty}dk^o \oint d^2\hat{k} T_{ab}   \left(\frac{\bar{\eth}\eth\big(\frac{l^al^b}{(k^cl_c)^2}\big)^+}{k^cl^+_c} + (-1)^l  \frac{\bar{\eth}\eth\big(\frac{l^al^b}{(k^cl_c)^2}\big)^-}{k^cl^-_c}\big)\right).
\end{equation}
    
The last term on the integral is a non vanishing form factor that yields the non trivial part of $\sigma^+_{l,J_l}(w)$. It does not depend on the stress energy tensor but it could give a vanishing contribution for some special cases, as for example, $T_{ab}=f(k)k_a k_b + g(k) \eta_{ab}$.

It is also useful to write down an equation relating $\sigma^+(u, \zeta,\bar{\zeta})$ with $\sigma^-(u, \zeta,\bar{\zeta})$ and the stress energy tensor $T_{ab}(x)$. To do this, we invert the frequency decomposition of equation (\ref{Zfinal}) and multiply by the corresponding $Y_{l,J}^{2}(\zeta,\bar{\zeta})$ obtaining

\begin{equation}\label{sigmafinal}
\sigma^+(u, \zeta,\bar{\zeta})+ \bar{\sigma}^-(u, \widehat{\zeta},\widehat{\bar{\zeta}})= i \int d^4k e^{i wu}T_{ab}(k) \oint d^2S' \left(\frac{\eth^2G_{00'}\bar{\eth'}\eth'\big(\frac{l'^al'^b}{(k^cl'_c)^2}\big)^+}{k^cl'^+_c} + \frac{\eth^2G_{\widehat{0}0'}\bar{\eth'}\eth'\big(\frac{l'^al'^b}{(k^cl'_c)^2}\big)^-}{k^cl'^-_c}\right),
\end{equation}

where $G_{\widehat{0}0'}=G_{00'}(\widehat{\zeta},\widehat{\bar{\zeta}},\zeta',\bar{\zeta}')$, $T_{ab}$ is written in spherical coordinates as $T_{ab}(k)=T_{ab}(k^o,w,\hat{k})$, and $w=\sqrt{\vec{k}.\vec{k}}$. Note that a generic $T_{ab}(x)$ does not satisfy the homogeneous wave equation, and thus, $k^a$ is not a null vector. As we can see in eq. (\ref{sigmafinal}), the coordinates $(\zeta,\bar{\zeta})$ and $(\hat{\zeta},\bar{\hat{\zeta}})$ are associated with the advanced and retarded solution respectively.
%Scalar field
\subsection{The scattering with a massless scalar field}

In this section we will apply the formula \cref{sigmafinal} to analyze the particular case when the matter field present is due to a scalar field. This implies the Lagrangian density from the field is
\begin{align*}
    \mathcal{L}=-\frac{1}{2}g^{ab}\partial_{a}\phi\partial_{b}\phi
\end{align*}
and the energy-momentum tensor
\begin{align}\label{ScalarFieldLagrangian}
    T^{ab}&=g^{ab}(-\frac{1}{2}\partial_c \phi \partial^c \phi)+ \partial^a \phi \partial^b \phi 
\end{align}
Contracting \cref{ScalarFieldLagrangian} with the null vectors $l^a$ , we obtain that the only term surviving the contraction is
\begin{align*}
   T_{ab}(x)l^a l^b=(l_a \partial^a \phi(x))^2.
\end{align*}
On the other hand, the scalar fields, which are solutions to the wave equation, can be written in terms of the Fourier transform
 \begin{equation}\label{PhiFourierTransform}
    \phi(y^{a})=\int  d^4k \left(e^{-i k^b y_b}a(k)+e^{i k^b y_b}\bar{a}(k) \right) \delta(k^ak_a)
\end{equation}
since the real scalar $\phi(y^{a})$ satisfies the wave equation. The energy-momentum tensor component that is needed for this calculation can be computed in momentum space as
\begin{align*}
    T_{ab}(k^c)l^a l^b&=\int d^4x e^{i k^b x_b}  (l^a \partial_a \phi(x^c))^2 
\end{align*}
then, 
$$T_{ab}(k^{a})l^a l^b=\int d^4k' a(k')a(k-k')  (l_a k'^{a}) l_b (k'^{b}-k^b)\delta(k'^ak'_a)\delta((k-k')^b(k-k')_b).$$
Note that both $k'^a$ and $(k-k')^b$ are null vectors whereas $k^b$ is not since $T_{ab}(x)l^a l^b$ does not satisfy the homogeneous wave equation, as one can see from a direct calculation.

The amplitude $a(k)$ in \cref{PhiFourierTransform} is directly related to the free data of the scalar field given at the boundary of the space time. One can thus directly compute the scattering between gravitational and scalar waves in terms of the different modes of the corresponding radiation fields at null infinity. To do that we first give the relationship between the Fourier transform $a(k')$ and the free data of the scalar field.

Assuming one has incoming scalar waves, one can write the solution of the wave equation as
\begin{equation}
    \phi(y^{a})=\oint  d^2S \dot{A}(v,\zeta,\bar{\zeta}),
\end{equation}
with $v=-y^{a} l^{-}_a$ and $A(v,\zeta,\bar{\zeta})$ the free data given at past null infinity. Furthermore, using a Fourier transform to obtain a frequency decomposition one has
\begin{equation}\label{Phinulldata}
    A(v,\zeta,\bar{\zeta})=\int_0^{\infty} dw \left(A(w,\zeta,\bar{\zeta})e^{-i w v}+\bar{A}(w,\zeta,\bar{\zeta})e^{i w v}\right).
\end{equation}
Thus,
\begin{equation}\label{Phinulldata}
    \phi(y^{a})=i\int_0^{\infty} w \; dw\oint  d^2S \left(A(w,\zeta,\bar{\zeta})e^{-i w v}-\bar{A}(w,\zeta,\bar{\zeta})e^{i w v}\right),
\end{equation}
which can be regarded as spherical coordinates in momentum space. One then has
\begin{equation}\label{Phinulldata}
    \phi(y^{a})=i\int  \frac{d^3k}{2w} 2i\left(\bar{A}(w,\zeta,\bar{\zeta})e^{-i k^a x_a}-A(w,\zeta,\bar{\zeta})e^{i k^a x_a}\right),
\end{equation}
where we have used $v=-l^{-a} x_a$ to obtain the above formula. Equation (\ref{PhiFourierTransform}) can be written as
\begin{equation}\label{Phinulldata}
    \phi(y^{a})=\int  \frac{d^3k}{2w} \left(a(k)e^{-i k^a x_a}+\bar{a}(k)e^{i k^a x_a}\right),
\end{equation}
by a straight calculation using $\delta(k^2)=  \delta((k^o -w)(k^o -w))$. It then follows that,
$$
a(w,\zeta,\bar{\zeta})=2i\bar{A}(w,\zeta,\bar{\zeta}),
$$
or
$$
a(w)_{lJ}=2i\bar{A}(w)_{lJ}.
$$

One can also show that using the positive frequency decomposition for $\phi$, together with $\delta(k^2)=  \frac{1}{2w_k}\delta((k^o -w))$ one obtains,
\begin{equation}\label{T(k)}
T_{ab}(k^{a})l^a l^b=\int \frac{d^3k_1}{4 w_1w_2} a(\vec{k}_1)a(\vec{k}- \vec{k}_1)  (l_a k_1^{a}) l_b (k_1^{b}-k^b)\delta(k^o -(w_1+w_2)),
\end{equation}
with $w_2=\sqrt{(\vec{k}- \vec{k}_1).(\vec{k}- \vec{k}_1)}$.

We can use the formula derived in the previous section to describe the scattering of incoming gravitational and scalar waves. For that, we give free radiation data for both fields and use formula \cref{sigmafinal} to obtain the outgoing gravitational waves at future null infinity represented by the radiation field $\sigma^+(u,\zeta,\bar{\zeta})$. In particular, we would like to compute the gravitational tail of the outgoing wave when both the incoming waves have compact support at past null infinity.

Selecting appropriate times $v_i$ and $v_f$ so that there are no incoming waves for $v<v_i$ or $v_f<v$, then we would like to obtain the gravitational radiation at future null infinity at sufficient large value of $u$ so that the "free" outgoing gravitational and scalar waves have died out. A Bondi time $u_f$ can be obtained such that a null plane that starts at $(v_f, \hat{\zeta}\hat{\bar{\zeta}})$, ends up at $(u_f,\zeta, \bar{\zeta})$. This plane is given by $y^a \ell^+_a= u_f=const.$ but it can also be written as $u_f=-y^a \hat{\ell}^-_a=v_f$. Thus, for times $u>u_f$, there are no "free"  outgoing waves, and directly from \cref{sigmafinal} one gets,

\begin{align}\label{sigmafi}
&\sigma^+(u, \zeta,\bar{\zeta}) = i \int d^{3}k_1 d^{3}k_2e^{-i wu}a^-(\vec{k}_1)a^-(\vec{k}_2)\oint d^2S' \frac{\eth^2G_{00'}}{(k_1+k_2)^{c}l'_c}\bar{\eth'}\eth'\big(\frac{k_{1a}k_{2b}l'^al'^b}{((k_1+k_2)^cl'_c)^2}\big),
\end{align}
with $w=\sqrt{(\vec{k}_1+ \vec{k}_2).(\vec{k}_1+ \vec{k}_2)}$ and we have used $l^{-a}=-\widehat{l}^{a}$, $G_{00'}= \widehat{G}_{00'}$ together with the coordinate transformation $\zeta \rightarrow \widehat{\zeta}$ to get the final form of eq. (\ref{sigmafi}). Note that $\sigma^-(u, \zeta,\bar{\zeta})$ is absent in the above equation since it has compact support and vanishes for $u= x^a l^+_a>u_f$. Likewise, $\phi(x)$ vanishes on the future null cone of $x^a$ and gives no contribution to the  the stress energy tensor on the integral on that cone. The corresponding $(l,I_l)$ mode for the positive frequency decomposition is given by 

\begin{align}\label{sigmafinal2}
&\sigma^+_{l,I_l}(w) = i \oint d^2\hat{k}\int d^{3}k_1 d^{3}k_2\mathcal{S}_{l,I_l}(
k,k_{1},k_{1})a^-(\vec{k}_1)a^-(\vec{k}_2),
\end{align}
with,
\begin{align}\label{smatrix}
\mathcal{S}_{l,I_l}(k,k_1,k_2)=\delta(w-(w_1+w_2)) \frac{Y_{l,I_l}^{0}(\hat{k})}{(k_1+k_2)^{c}l_c}\bar{\eth}\eth\big(\frac{k_{1}^ak_{2}^bl_al_b}{((k_1+k_2)^cl_c)^2}\big)
\end{align}
and $k^a=wl^a(\hat{k})$, $k_{1}^a=w_1l^a(\hat{k}_1)$, $k_{2}^a=w_2l^a(\hat{k}_2)$. Using a particle physics interpretation of the plane waves , the delta function can be interpreted as the conservation of energy of incoming and outgoing particles. Energy conservation, as opposed to full 4- momenta conservation, is a characteristic feature of the interaction with an external field, as one can also see in Section:8.7 of \cite{mandl2010quantum}.

It is worth making some remarks regarding these results. Eq. (\ref{sigmafi}) exhibits the tail of the gravitational wave due to the interaction with the scalar field since it is evaluated for times $u>u_f$. The equation for the modes relation (\ref{sigmafinal2}) is also interesting since it shows that one can obtain different harmonic components for the outgoing gravitational wave that may not be present in the incoming wave. Eq. (\ref{sigmafinal2}) is also interesting in a quantum field theory approach since it gives the probability scattering amplitude that an incoming graviton with a given value of quantum numbers end up with a different set of outgoing quantum numbers. In this case the scalar field is considered as an external classical field.

\subsubsection{Minimal coupling between the gravitational and a matter field}

Here we review the standard Lagrangian formulation for minimal coupling between matter and gravity since the two methods yield analogous results at a linearized level. However, as we will see below, there is a big difference between our formulation and the Green function approach based on the flat conformal structure. One can be extended in perturbation procedure, the other one cannot.

The Lagrangian for minimum coupling is given by
\begin{align}\label{lagrangian}
    \mathcal{L}=\mathcal{L}_G+\kappa_M \mathcal{L}_M
\end{align}
with 
\begin{align*}
\mathcal{L}_G= \sqrt{-g}R,
\end{align*}
$\kappa_M$ a suitable coupling constant, and $\mathcal{L}_M$ the lagrangian of the associated matter field. Writing down the Euler-Lagrange equations and then linearizing the obtained equations yields the first order set of equations from which a perturbation procedure can be implemented.

As a particular example we consider here a real massless scalar field $\phi$,

\begin{align}\label{lagrangian}
    \mathcal{L}_M =\frac{1}{2} \sqrt{-g}g^{ab} \partial_a \phi\partial_b \phi, 
\end{align}

It is straightforward to show that the trace free linearized perturbation of a flat metric obeys the wave equation with or without sources, i.e.,
\begin{align}
\Box h_{ab}= T_{ab} =\partial_a \phi\partial_b \phi - \frac{1}{2}g_{ab} \partial_c \phi\partial^c \phi,
\end{align}
with the massless field satisfying.
\begin{align}
\Box \phi=0.
\end{align}
Using the Green functions for the D'Alambertian operator in flat space one obtains 
\begin{align}\label{Z1plusb}
& g^{+}_{ab}=  \oint_{C^+_x} d^2S  \big(\dot{\sigma}^{+}(u,\zeta,\bar{\zeta})\bar{m}_a \bar{m}_b+ c.c.\big)  + \int_{N^-_x} d^4x'G^{+}(x,x')T_{ab}(x'),
\end{align}
\begin{align}\label{Z1plusb}
& g^{-}_{ab}=  \oint_{C^-_x} d^2S  \big(\dot{\sigma}^{-}(v,\zeta,\bar{\zeta})\bar{m}_a \bar{m}_b+ c.c.\big)  + \int_{N^-_x} d^4x'G^{-}(x,x')T_{ab}(x'),
\end{align}
with $G^{+}(x,x')$ ($G^{-}(x,x')$), the advanced (retarded) Green function of the wave operator. We thus obtain
a similar derivation to the one presented in the previous sections.

A natural question is then why should one follow the somewhat involved approach to obtain a standard result? The main reason is that there is no suitable perturbation calculation to obtain the higher order terms using the Green function of the flat D'Alambertian operator. If one tries to do so, one discovers that the null geodesics of the flat spacetime become either timelike or spacelike curves of the non trivial metric. Thus, they either end up at spacelike or at timelike infinity, i.e. never at null infinity and thus one cannot find a perturbation procedure that converges to the solution one is looking for.
Our formulation, on the other hand, is amenable to a perturbation approach that is well defined at every step of the perturbation scheme.
All the variables involved that were defined and used in the previous sections can be redefined at every step of the perturbation series since they have geometrical meanings, i.e, null cone cuts, affine parameters, Bondi coordinates, etc. For example, in the field equation for $Z$
\begin{align*}
\bar{\eth}^2 \eth^2 Z= \eth^2 \bar{\sigma}(Z,\zeta,\bar{\zeta})+ \bar{\eth}^2 \sigma(Z,\zeta,\bar{\zeta})+\int _Z^\infty\dot{\sigma}\dot{\bar{\sigma}}du-\int _s^\infty(\eth \bar{\eth}(\Omega^2)+g^{ab}\partial_a \Lambda\partial_b \bar{\Lambda})ds',
\end{align*}
all the variables involved can be obtained in a perturbation series leaving the geometrical meaning of the equation unchanged.

In addition, our formulation has a clear distinction between the true gravitational field i.e. the conformal structure, and the matter field. As one can see from the field equations, the function $\eth ^2 Z$ defines the conformal structure of the spacetime and only vanishes for a flat spacetime whereas the conformal factor $\Omega$ depends on the stress energy tensor.

\section{Summary and conclusions}\label{sectionVI}

In this work we use the  NSF approach to study the coupling of matter and gravity for the incoming/outgoing gravitational radiation given on the null boundaries of an asymptotically flat spacetime. 

We first derive a new field equation for the main variable since the original formulae contained some errors\cite{frittelli1995linearized}. This equation is a non-linear 4th order elliptic equation on the sphere with the matter term taking part in the inhomogeneity. Using an ab-initio assumption that the matter field preserves the regularity of the null cone cuts we solve this equation by a perturbation scheme in \cref{Zperturbation}.\\

We then consider the dispersion relation between incoming and outgoing gravitational waves when matter is present at the lowest non trivial level. This is done in Sec.\ref{sectionIII} by first obtaining a relation between the incoming $Z^-$ and outgoing $Z^+$  null foliations up to first order in the iterative process. This relation directly follows from  the uniqueness of the first order metric. \\
Formal solutions to the elliptic equation on $S^2$ are given for null and non-null matter fields. The former case yields, by means of \cref{ScatteringRelationC}, a relation for every mode in the decomposition of the gravitational shear $\sigma$. Although this relation gives the trivial scattering in every mode of the gravitational radiation in the absence of matter, it is important to obtain since it gives a non trivial relation between the incoming and outgoing radiation fields.\\
When matter is present, a similar relation can be derived for every mode. Despite being more difficult than in the non-matter case, this expression still gives a practical formula to work with. Indeed, the calculation of the last term in \cref{FinalNSF} can be directly implemented in numerical integration and get the desired mode of scattering. When the energy-momentum tensor $T_{ab}$ depends on the free data given on null infinity the calculations are greatly simplified. As an example, we consider a real, massless scalar field and write the mode relations for that particular case.\\
The above results could be thought of as the tree level approximation that one gets in field theory but using a different approach. A correlation with a more standard linear approximation using advanced or retarded Green functions of the flat wave equation is also given to emphasize the close relationship between the two approaches. In most scattering situations the tree level approximation yields the dominant terms of a perturbation series. 

However, there are several reasons why our formulation has several advantages over the standard approach. First, the perturbation approach based on null surfaces correctly incorporates the null free data for gravity given on past an future null infinities together with the conformal structure of the spacetime. The n-order term is a field given on a spacetime with null surfaces computed to the n-1 order. The Green function approach, on the other hand, always keeps the flat metric to define the null cone structure at each level of the approximation. Thus, these null lines of Minkowski space fail to reach either future or past null infinity and therefore never really reach the null free data and the perturbation procedure never converges to a solution of the problem. A second point is that the same free data in our formulation is used for any order of the perturbation procedure, i.e., the phase space is constructed once and for all in our approach. This is extremely important at a classical or quantum level since one can introduce either a canonical form with Poisson brackets or quantum commutation relations for fields given on the null boundaries that will not be modified as one proceeds with a perturbation calculation\cite{ashtekar1981asymptotic}. There is however an assumption that was used in this work that will have to be modified when higher order perturbations are considered. In our calculations we were assuming that the conformal completion of Minkowski spacetime is the Einstein universe, where spacelike infinity $i_o$ is a single point and  the conformal metric is regular everywhere. This assumption must be modified in the next orders of  perturbation calculations since they should include a non vanishing ADM mass. This in turns, forces to reexamine the notion of both the definition of $i_o$ and the regularity structure near spacelike infinity. A new mathematical treatment of $i_o$ has to be given to go along with higher order perturbation terms in order to map the kinematic structures of past and future null infinities. Although some progress has been made in this area\cite{dominguez1997phase, strominger2018lectures} there is still much work to be done to get a full understanding of the conformal completion of spacelike infinity for non trivial spacetimes\cite{friedrich1998gravitational,friedrich2000calculating}.

\begin{acknowledgments}
This research has been supported by grants from CONICET and the Agencia Nacional de Ciencia y Tecnolog\'ia.
\end{acknowledgments}

\nocite{*}

\bibliography{main}

\appendix

\section{Derivation of $\sigma$ modes relation}\label{AppendixA}

We first derive the relationship for the "free" part of Z, that satisfies the wave equation, since it corresponds to undisturbed gravitational waves that propagate from minus to plus null infinity. $Z^{+}_{l,I_l}$ is given by

\begin{align}\label{partialZ1plus}
& Z^{+}_{l,I_l}=  \oint  \big(Y_{l,I_l}^{-2}(\zeta,\bar{\zeta})  \sigma^{+}(u,\zeta,\bar{\zeta}) +Y_{l,I_l}^{2}(\zeta,\bar{\zeta}) \bar{\sigma}^{+}(u,\zeta,\bar{\zeta}) \big) d^2S,
\end{align}
and we assume the free data $\sigma^{+}(u,\zeta,\bar{\zeta})$ admits a positive frequency decomposition,
$$
\sigma^{+}(u,\zeta,\bar{\zeta})= \int_0^\infty w^2 \; dw \sigma^{+}(w,\zeta,\bar{\zeta}) e^{-iwu}.
$$
We then rewrite the free data part of  $Z$ as
\begin{equation}
    Z^{+}_{l,I_l} = \int_{0}^{\infty} w^2 \; dw \oint d^2 S[Y^{-2}_{l,I_l}(\zeta,\bar{\zeta}) \sigma^{+}(w, \zeta,\bar{\zeta}) e^{-iw x^al_a} + Y^{2}_{l,I_l}(\zeta,\bar{\zeta}) \bar{\sigma}^{+}(w, \zeta,\bar{\zeta}) e^{iw x^al_a} ],
\end{equation}
which can be rewritten as
\begin{equation}\label{fourier Z}
    Z^{+}_{l,I_l} = \int d^3k[Y^{-2}_{l,I_l}(\hat{k}) \sigma^{+}(k) e^{-i x^ak_a} + Y^{2}_{l,I_l}(\hat{k}) \bar{\sigma}^{+}(k) e^{i x^ak_a} ],
\end{equation}
using a spherical decomposition of the 3-dim momentum space. In the above equation the coordinates $(\zeta,\bar{\zeta})$ on the sphere have been rewritten as $(\hat{k})$, and $k^a = w l^a(\hat{k})$. Note that $k^a k_a=0$, and we have a field that satisfies the wave equation. Using the eigenfunctions $e^{i x^ak_a}$ we obtain the Fourier transform of the above equation as

\begin{equation}\label{Z-free}
\int d^3 x e^{i x^ak_a}\overleftrightarrow{\partial_t}Z^{+}_{l,I_l} = Y^{-2}_{l,I_l}(\hat{k}) \sigma^{+}(w,\hat{k})\ .
\end{equation}
The operator $\overleftrightarrow{\partial_t}$ is commonly used in Quantum field theory and given two scalars $a$ and $b$, it is defined as $a\overleftrightarrow{\partial_t}b=a(\partial_tb)-(\partial_t a)b$.
Finally, integrating on the sphere \cref{Z-free} yields
\begin{equation}\label{Zlm}
    \oint d^2 \hat{k}\int d^3 x e^{i x^ak_a}\overleftrightarrow{\partial_t}Z^{+}_{l,I_l}=\sigma^{+}(w)_{l,I_l}.
\end{equation}

$\widehat{Z^{-}}_{l,I_l}$, on the other hand, is given by

\begin{align}\label{Zminus}
& \widehat{Z^{-}}_{l,I_l}= (-1)^l \oint  \big(Y_{l,I_l}^{-2}(\zeta,\bar{\zeta})  \sigma^{-}(v,\zeta,\bar{\zeta}) +Y_{l,I_l}^{2}(\zeta,\bar{\zeta}) \bar{\sigma}^{-}(v,\zeta,\bar{\zeta}) \big) d^2S,
\end{align}
thus, to compare both expressions one must first rewrite eq.(\ref{Zminus}) as a function of $u=x^al^+_a$ instead of $v= -x^al^-_a$. From $ l^-_a=-\widehat{l^+}_a$ we obtain $v=\widehat{u}$. Performing a change of variables $\zeta \rightarrow \widehat{\zeta}$ in Eq.(\ref{Zminus}) we obtain
\begin{align}
& \widehat{Z^{-}}_{l,I_l}= \oint  \big(Y_{l,I_l}^{-2}(\zeta,\bar{\zeta})\bar{\sigma}^{-}(u,\widehat{\zeta},\widehat{\bar{\zeta}})   +Y_{l,I_l}^{2}(\zeta,\bar{\zeta})\sigma^{-}(u,\widehat{\zeta},\widehat{\bar{\zeta}})  \big) d^2S,
\end{align}
where we have used $Y_{l,I_l}^{2}(\widehat{\zeta},\widehat{\bar{\zeta}})= (-1)^lY_{l,I_l}^{-2}(\zeta,\bar{\zeta})$. Note that $\bar{\sigma}^{-}(u,\widehat{\zeta},\widehat{\bar{\zeta}})$ is a s.w 2 function and plays the same role as $\sigma^{+}(u,\zeta,\bar{\zeta})$ in $Z^{+}_{l,I_l}$. Thus, 
\begin{equation}
    \widehat{Z^{-}}_{l,I_l} = \int_{0}^{\infty} w^2 \; dw \oint d^2 S[Y^{-2}_{l,I_l}(\zeta,\bar{\zeta}) \bar{\sigma}^{-}(w,\widehat{\zeta},\widehat{\bar{\zeta}}) e^{-iw x^al_a} + Y^{2}_{l,I_l}(\zeta,\bar{\zeta}) \sigma^{-}(w,\widehat{\zeta},\widehat{\bar{\zeta}}) e^{iw x^al_a} ],
\end{equation}
Finally, 
\begin{equation}\label{Zlm}
    \oint d^2 \hat{k}\int d^3 x e^{i x^ak_a}\overleftrightarrow{\partial_t}\widehat{Z^{-}}_{l,I_l}=(-1)^l\bar{\sigma}^{-}(w)_{l,I_l},
\end{equation}
and from \cref{ScatteringRelationC} we get
\begin{equation}\label{Zplus +Zminus}
\sigma^{+}(w)_{l,I_l}+(-1)^l\bar{\sigma}^{-}(w)_{l,I_l}=0.
\end{equation}

We now address the mode decomposition of the full term either at future or past null infinity including the integration on the future or past null cones from the point $x^a$ starting with the advanced solution,  i.e.,
\begin{align}\label{Z1plus}
& Z^{+}_{l,I_l}=  \oint  \big(Y_{l,I_l}^{-2}  \sigma^{+}(u,\zeta,\bar{\zeta}) + c.c.\big)+  Y_{l,I_l}^{0}\int_0^\infty ds \bar{\eth} \eth g_{01}(y^c,\zeta,\bar{\zeta})\big) d^2S,
\end{align}
where the last term is integrated on the future null cone from $x^a$ and $c.c$ stands for the complex conjugate term. It is useful to have a spherical harmonic decomposition before the integral on the null cone is performed. for this we expand $T_{ab}(y^a)$ in a Fourier decomposition,
\begin{equation}\label{FFT}
    T_{ab}(y^{c})=\int d^4k T_{ab}(k^c) e^{-i k^a y_a}= \int d^4k \bar{T}_{ab}(k^c) e^{i k^a y_a},
\end{equation}
where we have used the real condition on the last equation. We now write

$$
\bar{\eth} \eth g_{01}(y^{+c},\zeta,\bar{\zeta})= T_{ab}(y^{+c})  \bar{\eth}\eth\left(\frac{l^{+a}l^{+b}}{(k^cl^{+}_c)^2}\right)
$$
where $y^{+c}=x^c+ sl^{+c}$. Inserting (\ref{FFT}) into (\ref{Z1plus}) yields,
$$
\int_0^\infty ds\oint  d^2S Y_{l,I_l}^{0} \bar{\eth} \eth g_{01}(y^{+c},\zeta,\bar{\zeta}) =
i \int d^4k e^{-i k^a x_a}T_{ab}(k^c)  \oint  Y_{l,I_l}^{0} \bar{\eth} \eth\left(\frac{l^{+a}l^{+b}}{(k^cl^{+}_c)^2}\right)\frac{d^2S}{k^cl_c},
$$
To compute the value of $\sigma^+_{l,I_l}(w)$ at future null infinity, and calling $Z_{free}$ to part of $Z$ without the matter field term, it is convenient to have a 4-dim Fourier decomposition of $Z_{free}(x^a)_{l,I_l}$ as,

\begin{equation}\label{fourierZ2}
    Z_{free}(x^a)_{lJ_l} = \int d^4k\delta(k^ak_a)\theta(k^o)[Y^{-2}_{l,I_l}(\hat{k}) \sigma(w_k,\hat{k}) e^{-i x^ak_a} + Y^{2}_{l,I_l}(\hat{k}) \bar{\sigma}(w_k,\hat{k}) e^{i x^ak_a} ].
\end{equation}
Taking the Fourier transform of the above equation gives, 
\begin{equation}
    \int d^4x e^{i kx} Z_{free}(x^a)_{lJ_l} = \frac{\delta(k^o - w_k)}{2 w_k} Y^{-2}_{l,I_l}(\hat{k}) \sigma(w_k,\hat{k}).
\end{equation}
Using  spherical coordinates on momentum space, integrating on the frequency and on the sphere in momentum space gives,
\begin{equation}
    \int_o^\infty dk^o \oint d^2\hat{k} \frac{\delta(k^o - w_k)}{2 w_k} Y^{-2}_{l,I_l}(\hat{k}) \sigma(w_k,\hat{k}) =  \frac{1}{2 w_k}\sigma(w_k)_{l,J_l} .
\end{equation}
Finally, starting with  (\ref{Z1plus}), and following the same steps as before yields
\begin{equation}\label{Z+lm}
    Z^+_{l,I_l}(w)=\sigma^+_{l,I_l}(w) + i \int_0^{\infty}dk^o \oint d^2\hat{k} T^+_{ab}(k^o,w\hat{k}^i) \oint Y_{l,I_l}^{0} \bar{\eth} \eth\left(\frac{l^al^b}{(k^cl_c)^2}\right)\frac{1}{k^cl_c},
\end{equation}
with $k^a=(k^o,w\hat{k}^i)$.

To obtain the contribution of the retarded solution we first write
\begin{align}\label{Z1-}
     Z^{-}(x^a ,\zeta,\bar{\zeta})= \oint_{S^2} \left(\bar{\eth}^{\prime 2}G^-_{00}(\zeta,\bar{\zeta},\zeta',\bar{\zeta}') \sigma^{-}(v,\zeta',\bar{\zeta}')+ c.c.\right) dS^{\prime},
 \end{align}
with $v= -x^a l^{-}_a= -x^a \hat{l}^{+}_a$. In order to compare with $Z^{-}(x^a ,\zeta,\bar{\zeta})$ we perform a change of variables $\zeta',\bar{\zeta}' \rightarrow \widehat{\zeta}',\widehat{\bar{\zeta}}'$ giving

\begin{align}
Z^{-}_{free}(x^a ,\widehat{\zeta},\widehat{\bar{\zeta}})= \oint_{S^2} \left(\bar{\eth}^{\prime 2}G^-_{00}(\widehat{\zeta},\widehat{\bar{\zeta}},\widehat{\zeta}',\widehat{\bar{\zeta}}') \sigma^{-}(-u,\widehat{\zeta}',\widehat{\bar{\zeta}}')+ c.c.\right) dS^{\prime}.
 \end{align}
 
Thus, the positive frequency decomposition of $Z^{-}(x^a ,\widehat{\zeta},\widehat{\bar{\zeta}})$ is given by
 
 \begin{align}
Z^{-}_{free}(x^a ,\widehat{\zeta},\widehat{\bar{\zeta}})= \int_0^\infty dw \oint_{S^2} dS^{\prime}\left(\bar{\eth}^{\prime 2}G^-_{00}(\widehat{\zeta},\widehat{\bar{\zeta}},\widehat{\zeta}',\widehat{\bar{\zeta}}') \sigma^{-}(w,\widehat{\zeta}',\widehat{\bar{\zeta}}')e^{iw x^al'_a}+ c.c.\right),
 \end{align}
 and,
 
 \begin{equation}
    Z^{-}_{free}(x^a)_{l,I_l} = (-1)^l \int_{0}^{\infty} dw\oint d^2 S'[Y'^{2}_{l,I_l}(\zeta',\bar{\zeta}') \bar{\sigma}^{-}(w, \zeta',\bar{\zeta}') e^{-iw x^al'_a} +  Y'^{-2}_{lJ_l}(\zeta',\bar{\zeta}') \sigma^{-}(w, \zeta',\bar{\zeta}') e^{iw x^al'_a}],
    \end{equation}
 where we have used $\widehat{Y^2}_{l,I_l}= (-1)^l Y^{-2}_{l,I_l}$ and 
 $\widehat{Y^2_{l,I_l}Y^{-2}_{l',I_l'}}= Y^2_{l,I_l}Y^{-2}_{l',I_l'}$. Thus, directly from 
 $$
 Z^{+}_{free}(x^a ,\zeta,\bar{\zeta})+Z^{-}_{free}(x^a ,\widehat{\zeta},\widehat{\bar{\zeta}})=0,
 $$
 we get
 $$
 \sigma^{+}_{l,I_l}(w) + (-1)^l\bar{\sigma}^{-}_{l,I_l}(w)=0.
 $$
 
 To obtain the past null cone contribution to (\ref{Z+lm}) we have to replace $y^{+c}=x^c+ sl^{+c}$ by $y^{-c}=x^c+ sl^{-c}$,
 $$
\int_0^\infty ds\oint  d^2S Y_{l,I}^{0} \bar{\eth} \eth g_{01}(y^{c-},\zeta,\bar{\zeta}) =
i \int d^4k e^{-i k^a x_a}T_{ab}(k^c)  \oint  Y_{l,J_l}^{0} \bar{\eth} \eth\left(\frac{l^{-a}l^{-b}}{(k^cl^{-}_c)^2}\right)\frac{d^2S}{k^cl^{-}_c}.
$$
 Thus,
 \begin{equation}\label{Z-lm}
    (-1)^lZ^-_{l,I_l}(w)=\bar{\sigma}^-_{l,I_l}(w) - i \int_0^{\infty}dk^o \oint d^2\hat{k}  T_{ab}(k^c)\oint Y_{l,I_l}^{0} \bar{\eth} \eth\left(\frac{l^al^b}{(k^cl_c)^2}\right)^-\frac{1}{k^cl^-_c},
\end{equation}
where the relative minus sign arises from the odd number of $l_c$ terms in the stress energy tensor term.

Finally, from $Z^{+}(x^a ,\zeta,\bar{\zeta})+Z^{-}(x^a ,\widehat{\zeta},\widehat{\bar{\zeta}})=0$, we get

\begin{equation}\label{Zfinal}
  \sigma^+(w)_{l,I_l} + (-1)^l\bar{\sigma}^{-}(w)_{l,I_l} + i \oint d^2S Y_{l,I_l}^{0}\int_0^{\infty}dk^o \oint d^2\hat{k} T_{ab}   \left(\frac{\bar{\eth}\eth\big(\frac{l^al^b}{(k^cl_c)^2}\big)^+}{k^cl^+_c} + (-1)^l  \frac{\bar{\eth}\eth\big(\frac{l^al^b}{(k^cl_c)^2}\big)^-}{k^cl^-_c}\big)\right)=0.
\end{equation}

\end{document}